\documentclass[12pt]{article}

\input epsf
\usepackage{amssymb,wrapfig,amsmath}
\usepackage{xypic,multirow}
\usepackage[matrix,arrow,curve]{xy}
\usepackage{epsfig}

\begin{document}
\addtolength{\baselineskip}{.5mm}
\newlength{\extraspace}
\setlength{\extraspace}{1.5mm}
\newlength{\extraspaces}
\setlength{\extraspaces}{2mm}

\def\nn{{\cal N}}
\def\rr {{\Bbb R}}
\def\cc {{\Bbb C}}
\def\C {{\mathbb C}}
\def\pp {{\Bbb P}}
\def\zz {{\Bbb Z}}
\def\del {\partial}
\def\cy {Calabi--Yau}
\def\ka {K\"ahler}
\def\be{\begin{equation}}
\def\ee{\end{equation}}
\def\vp{\varphi}
\def\oo{\mathcal{O}}
\newcommand{\inv}[1]{{#1}^{-1}} 
\newcommand{\Section}[1]{\section{#1} \setcounter{equation}{0}}
\makeatletter
\@addtoreset{equation}{section}
\makeatother
\renewcommand{\theequation}{\thesection.\arabic{equation}}

\begin{center}
{\hfill RUNHETC-2009-14\\}

\medskip

{\Large \bf{BPS State Counting in Local Obstructed Curves from Quiver Theory and Seiberg Duality}}

\bigskip

Wu-yen Chuang and Guang Pan
\medskip

{\it NHETC, Department of Physics, Rutgers University}

{\it 126 Frelinghuysen Rd, New Jersey 08854, USA}

{\it wychuang@physics.rutgers.edu, guangpan@physics.rutgers.edu}

\bigskip

{\bf Abstract}
\end{center}

In this paper we study the BPS state counting in the geometry of
local obstructed curve with normal bundle $\mathcal{O} \oplus
\mathcal{O}(-2)$. We find that the BPS states have a framed quiver
description. Using this quiver description along with the Seiberg
duality and the localization techniques, we can compute the BPS state indices in different chambers
dictated by stability parameter assignments. This provides a well-defined
method to compute the generalized Donaldson-Thomas invariants. This method 
can be generalized to other affine ADE quiver theories.

\tableofcontents

\medskip \medskip

\section{Introduction}

Recently there has been much progress in the D-branes BPS state
index counting and its wall crossing behavior across the lines of marginal
stabilities \cite{DM}. Physically the change in the BPS state index can be
understood from the fact that in supergravity the separation of the multiple centered
solution goes to infinity when the moduli approach the marginal stability
wall \cite{Denef1}.  This infinite separation causes the bound state to leave the
Hilbert space and results in the discrete jumps in the BPS state spectrum.

On the other hand, we believe these BPS indices correspond to the
generalized Donaldson-Thomas invariant defined in mathematical literatures \cite{KS}\cite{Joyce}.
In the generalized Donaldson-Thomas theory, the integration of the virtual fundamental
class over the moduli space of the ideal sheaves is replaced by counting the stable objects
in the derived category.

The physical interpretation of the wall crossing formula was given if we consider
the BPS instanton corrections to the hyperk\"ahler metric of the moduli space of 4d $N=2$ theory
reduced on a circle \cite{GMN}. The wall crossing formula is equivalent to the statement that
this hyperk\"ahler metric is continuous with respect to the moduli variation.

In \cite{Szendroi} Szendr\H{o}i considered the moduli space of the framed cyclic modules in the conifold
path algebra. The torus fixed points in the moduli space are in one to one correspondence with
the torus fixed ideals in the path algebra. With the knowledge that all the torus fixed ideal are
generated by monomials, one could classify all the fixed points and arrange them into
\textit{pyramid partitions} of length 1 \textit{empty room configuration} (ERC).
These configurations can be summed exactly\cite{BenYoung}. To find out the physical
interpretation of Szendr\H{o}i's partition function, the authors in \cite{JM} used supergravity
techniques and obtained the BPS spectra on a resolved conifold in all different chambers.
In particular, the result in \cite{Szendroi} was reproduced as a BPS state partition
function in certain K\"ahler chamber. Later the authors in \cite{CJ} found that the
partition function of the pyramid partition with different ERC
actually correspond to BPS state partition function in different chambers. They
made conjecture about a new finite type pyramid partitions, which will asymptote to the stable
pair invariant (aka Pandharipande-Thomas or PT invariant) \cite{PT} when the length of the
ERC goes to infinity. The conjecture was later proven by \cite{Naka}.

This relation between the different ERCs and K\"ahler chambers arises if we start from the original
framed conifold quiver and then perform the Seiberg duality to bring back the stability
parameters (FI parameters) to the cyclic FI parameter assignment.\footnote{By "cyclic
FI parameter assignment" we mean the FI parameter of the framing node is positive and
the other FI parameters are negative.} This mutation, or Seiberg duality in physics term,
will result in the change in the framing data and superpotential. From new framing data and
new superpotential, one determines the new ERC for the pyramid partitions.

The Seiberg duality of the quiver theory sometimes simply provides us different aspects of the same thing. In this case, however,
after the mutation it will become extremely easy to visualize the arrangement of the torus fixed
points in the cyclic chamber of the mutated quiver with new framing data and new superpotential.

In this paper we consider a local obstructed
$\mathbb{P}^1$ with normal bundle $\mathcal{O} \oplus \mathcal{O}(-2)$.
The quiver theory with superpotential was given in \cite{AK}. The superpotential
deformation is determined by the obstruction of the $\mathbb{P}^1$. In order to
study the D6-D2-D0 system on this geometry we introduce a framing node into
the quiver. We find that \textit{the mutated quiver in the cyclic chamber computes
what the original framed quiver in certain chamber does.} In other words, due to the
simplicity of the cyclicity, we can bring the quiver theory
back to the cyclic chamber by using Seiberg duality, which
will change the new superpotential and the arrow structure connecting the framing node.

We then consider the framed cyclic $\mathcal{A}$-module of the mutated quiver, where
$\mathcal{A}$ is the path algebra of the mutated quiver.
In this quiver theory there are still $(\mathbb{C}^*)^2$ actions leaving F-term relations invariant,
although the geometry is nontoric.
Once again the torus fixed point in the moduli
space of the framed cyclic $\mathcal{A}$-module should be in one-to-one correspondence with
the torus fixed ideals of the path algebra. These ideals are generated by monomials
and can be arranged into \textit{filtered pyramid partitions of $(n,k)$ (in)finite ERC}\footnote{We will explain (n,k) (in)finite ERC in Section \ref{sec:class}.}, where $n$
is the degree of the obstruction in the geometry and $k$ is the number of the framing arrows.

After the classification we can utilize the localization techniques
to compute the contribution from each fixed point.
Unlike conifold in which the $(\mathbb{C}^*)^2$ actions preserve the superpotential, the deformation
complex is not self-dual in this case. Therefore the local contribution of each fixed point will be
generically a rational function in terms of localization parameters.  After summing all the fixed
points for a given dimension vector assignment we find the integrability is recovered.

Since the local obstructed $\mathbb{P}^1$ geometries can be deformed into $n$ isolated conifold
geometries by using complex deformation and every
$\mathbb{P}^1$ is homologous to each other, we expect the localization result should reproduce $n$ copies
of the conifold. We compare our result with this expectation and find perfect match.

The paper is organized as follows. In section \ref{sec:geometry} we introduce the geometry on which we will
study the BPS state counting. Section \ref{sec:framed} is about the Seiberg duality in affine $A_1$ quiver
and what this mutated framed quiver computes. In section \ref{sec:class}  we classify all the torus fixed points
in the mutated framed affine $A_1$ quiver by looking for toric fixed annihilator ideals.
In section \ref{sec:melt} we comment on the implication for the melting crystal model.
In section \ref{sec:conclusion} we conclude and discuss some possible future directions. In appendix 
\ref{append:A}\ref{append:B}\ref{append:C}\ref{append:D} 
we give a short review on the mutation and present the new result about the mutation
on a quiver theory with framing and adjoint fields. Appendix \ref{append:E} contains the explicit construction
of the deformation complex and useful information for computing the local contributions of the torus fixed points.

\section{The Geometry: A $\mathbb{P}^1$ with Obstruction}
\label{sec:geometry}
The main geometry discussed in this paper is a local obstructed $\mathbb{P}^1$.
Such local $\mathbb{P}^1$ can be described explicitly in patches with transition functions \cite{AK} \cite{Zhou}:
\begin{eqnarray}
\label{eq:geo}
w &=& x^{-1} \nonumber \\
z_1 &=& x^2 y_1 + x y_2^n \nonumber \\
z_2 &=& y_2
\end{eqnarray}
where $(n+1)$ is the degree of the obstruction.

Define a map from these two patches $(w,z_1,z_2)$ and $(x,y_1,y_2)$ to $\mathbb{C}^4$:
\begin{eqnarray}
v_1=z_2=y_2, \ \ \ v_2=z_1=x^2y_1+ xy_2^n, \nonumber \\
v_3=wz_1=xy_1+y_2^n, \ \ \ v_4=w^2 z_1 -wz_2^n=y_1.
\end{eqnarray}

Then the geometry is defined by $v_2v_4-v_3^2+v_3 v_1^n=0$ in $\mathbb{C}^4$, or by a change of
coordiantes, $u_1^2+u_2^2+u_3^2+u_4^{2n}=0$ in $\mathbb{C}^4$.

The quiver theory can be obtained by performing an Ext group computation
and using the $A_{\infty}$ structure \cite{AK}.
\begin{equation}
\label{eq:q}
\xymatrix{
\bullet 1 \ar@(ul,dl)[]|{C_1}  \ar@/^/@{->>}[rr]^{A_1, A_2} && \bullet 2  \ar@(ur,dr)[]|{C_2} \ar@/^/@{->>}[ll]^{B_1, B_2}  \\
}
\end{equation}

The superpotential is,
\begin{equation}
W= w(C_1)-w(C_2) + C_1 ( B_1 A_1 + B_2 A_2) -C_2 ( A_1 B_1 +A_2 B_2),
\end{equation}
where $w(C_i)= \frac{1}{n+1}C_i^{n+1}$.

Taking partial derivatives of $W$ gives the following relations: 
\begin{eqnarray} 
\label{rel1}
C_1^n+B_1A_1+B_2A_2=0 \nonumber \\
C_2^n+A_1B_1+A_2B_2=0 \nonumber \\
C_1B_i=B_iC_2 \nonumber \\
C_2A_i=A_iC_1. 
\end{eqnarray}

First note that we get the same relation as $n=1$ (Klebanov-Witten) by
commuting $C_i$ k times through $A_i$ and $B_i$

\begin{eqnarray} 
\label{eq:commuteC}
A_iC_1^k&=C_2^kA_i \nonumber \\
B_iC_2^k&=C_1^kB_i
\end{eqnarray}

and plugging the relation (\ref{eq:commuteC}) back into (\ref{rel1}):
\begin{eqnarray}
\label{eq:ABA}
A_1B_iA_2&=A_2B_iA_1 \nonumber \\
B_1A_iB_2&=B_2A_iB_1
\end{eqnarray}

One may identify the center of the algebra by:
\be
\begin{split}x&=A_1B_1+B_1A_1-A_2B_2-B_2A_2\\y&=2(A_1B_2+B_2A_1)\\z&=2(A_2B_1+B_1A_2)\\u&=C_1+C_2
\end{split}
\ee
which satisfies: \be x^2+yz-u^{2n}=0\ee

Namely the center of the path algebra of the quiver (\ref{eq:q}) is the same geometry we are looking at. For
a generic superpotential deformation 
$w(C_i) = \frac{1}{n+1}C_i^{n+1} + a_{n} C_i^{n} +\cdots$, we can repeat this and obtain the 
center of the path algebra as $x^2+yz-(w'(u))^2=0$.\footnote{The facts that the center of the algebra provides the Calabi-Yau geometry that one is probing
and the 4d quiver gauge theory vacua correspond to representations of noncommutative algebras were first pointed out in \cite{Berenstein1}\cite{Berenstein2}.
The quiver theory in this paper is the quiver quantum mechanics describing the BPS particles \cite{denefhalo}. The representations of the quiver theory then correspond to 
the BPS bound states.}

Recall that each node of the quiver represents an object in the derived
category of coherent sheaves. In this case the objects are
$\mathcal{O}_C$ and $\mathcal{O}_C(-1)[1]$
respectively.\footnote{Later in the paper we will have to use
$\mathcal{O}_C(-k)[1]$ ($\bar{D2} + (k-1)D0$) and
$\mathcal{O}_C(-k-1)$ ($D2 + k \bar{D0}$) as the basis of the
quiver.} This quiver is also called as unframed affine $A_1$ quiver
with degree $n+1$ superpotential deformation. In the next section we
will add a framing node (D6 node) into this quiver theory and impose
the stability condition.

\section{Framed Affine $A_1$ Quiver with Superpotential Deformation}
\label{sec:framed}

Since our goal is to count D6-D2-D0 bound states in the local geometry (\ref{eq:geo}),  we need to introduce a new node for
$D6$ brane wrapping the whole geometry and then impose certain
stability condition. In the quiver theory the stability condition is imposed by assigning FI parameters to each node
in the quiver. Counting BPS states is equivalent to counting the stable representation of
the quiver theory under this FI parameter
assignment \cite{King}. A commonly used stability condition is to require the representations of the
quiver theory to be cyclic. That is to say, every representation is generated by a vector field $\mathbb{C}$ in the vector space $V_1$ at node 1.

The added node for $D6$ brane represents the structure sheaf $\mathcal{O}_X$ and the resulting quiver theory is
as follows.

\begin{equation}
\xymatrix{
& \bullet 0  \ar[ddl]_{Q_1} &  \\
& & \\
\bullet 1 \ar@(ul,dl)[]|{\tilde{C}_1} \ar@/^/@{->>}[rr]^{A_1, A_2} && \bullet 2
\ar@(ur,dr)[]|{C_2} \ar@/^/@{->>}[ll]^{B_1, B_2}  \\
}
\end{equation}

By some simple argument we can show that
cyclicity implies that the FI parameter $\theta_0$ for the D6 node is positive which other FI parameters are negative \cite{CJ}.
Note that, if the degree $n=1$, we can integrate out $C_1$ and $C_2$ in the superpotential and reproduce the conifold
quiver \cite{KW}.
Such a conifold quiver in the cyclic chamber will compute the \textit{noncommutative Donaldson-Thomas invariant} a l\`a Szendr\H{o}i.
In Szendr\H{o}i partition function the MacMahon factor is present, which signals the noncompactness of the moduli space.
The physical interpretation is that in general the D0-branes do not have to be bound to the D2-branes.

In fact we would like to find a chamber, dictated by the assignment of the $\theta$s, describing the PT chamber.
The PT invariants are supposed to be the same as DT or GW invariant, modulo the MacMahon.
Such a chamber in conifold quiver was actually found in \cite{CJ}\cite{Naka}. In order to get to such a chamber, we have to
start from a different quiver theory with a reversed framing arrow in the cyclic chamber and still keep $\theta_0$
positive.\footnote{This procedure is NOT the same as taking the \textit{dual} representation of the quiver because we did
not flip the sign of the $\theta$s.}
The cyclic chamber for this reversed framed quiver is actually an empty chamber, in which every representation is unstable.\footnote
{It can be easiy shown by making an unstable
subrepresentation for every representation.}

\medskip \medskip

{\bf Relevant digression (Conifold example) :} Since we will borrow some of the conifold results, we give a brief review on it now. The framed conifold quiver
in NCDT chamber is given by the following quiver with the usual quartic superpotential and $\theta_0>0$, $\theta_1, \theta_2 <0$:

\begin{equation}
\xymatrix{
& \bullet 0  \ar[ddl]_{Q_1} & \theta_0 >0 \\
& & \\
\bullet 1 \ar@/^/@{->>}[rr]^{A_1, A_2} && \bullet 2 \ar@/^/@{->>}[ll]^{B_1, B_2}  \\
}
\end{equation}
\be
W=A_1B_1A_2B_2 - A_1B_2A_2B_1
\ee

If we keep the same $\theta$ parameter assignment and reverse $Q_1$, we obtain a quiver theory
of which the generating function is simply 1, due to the fact mentioned in the footnote.
\begin{equation}
\xymatrix{
& \bullet 0  & \theta_0 >0 \\
& & \\
\bullet 1 \ar[uur]^{Q_1}  \ar@/^/@{->>}[rr]^{A_1, A_2} && \bullet 2
\ar@/^/@{->>}[ll]^{B_1, B_2}  \\
}
\end{equation}

\begin{figure}[ht]
\centering
\epsfig{file=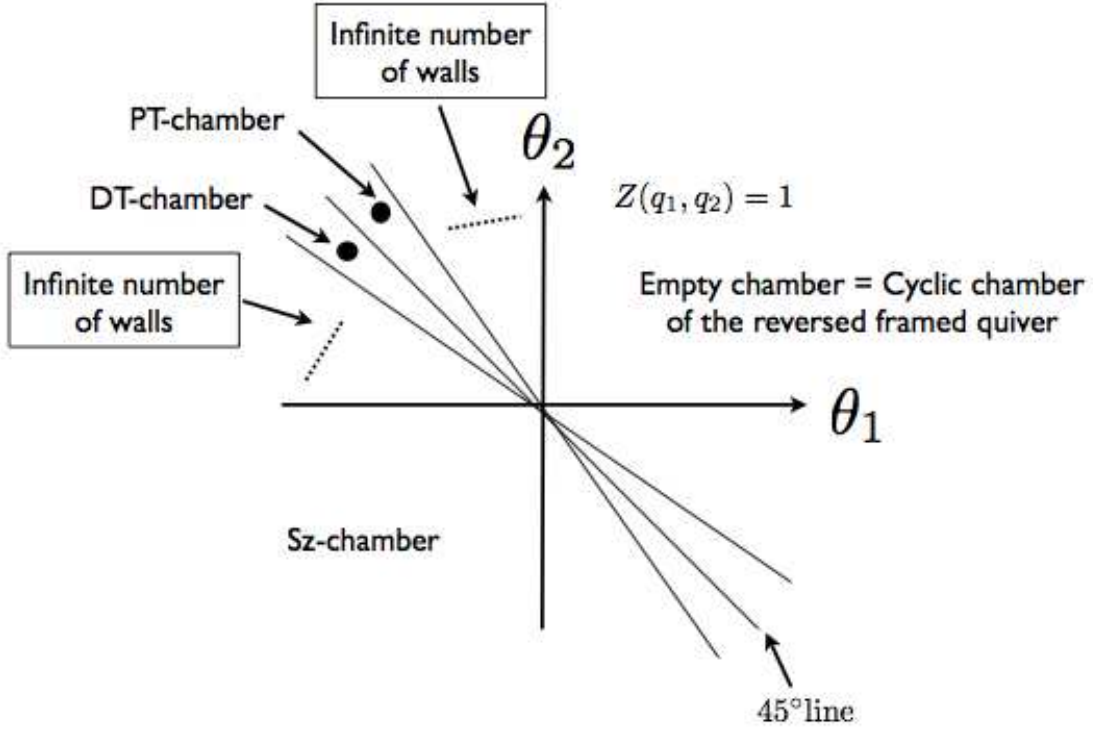, width=15cm}
\caption{This is the chamber structure of the original conifold quiver with only one
framing arrow.}
\label{fig:theta}
\end{figure}

The PT chamber is actually described by this quiver theory with $\theta_1$ and $\theta_2$
being very close to the 45 degree line in the $\theta_1$-$\theta_2$ plane \cite{Naka}. See
Fig.\ref{fig:theta} for details.

However we can use the Seiberg duality to get a mutated quiver with
cyclic $\theta$ assignment. In the mutated quiver the number of
framing arrows will increase to number $k$ as illustrated below.

\begin{equation}
\xymatrix{
& \bullet 0  \ar[ddl]_{Q_1,...,Q_k} &  \\
& & \\
\bullet 1 \ar@/^/@{->>}[rr]^{A_1, A_2} && \bullet 2 \ar[uul]_{P_1,...,P_{k+1}}\ar@/^/@{->>}[ll]^{B_1, B_2}  \\
}
\end{equation}

Under the mutation at node 2, the $\theta$ parameters will change in the following way\cite{Nagao}:
\be 
\label{eq:FI}
\theta_1 \rightarrow \theta_1 + 2 \theta_2, \ \ \theta_2 \rightarrow - \theta_2.
\ee
After alternating $k$ steps of the mutations, the $\theta$ parameters of the 
mutated quiver $(\tilde{\theta}_1, \tilde{\theta}_2)$ is related to the $\theta$ parameters
of the original quiver as follows:
\be 
\tilde{\theta}_1 =  k \theta_1 + (k+1)\theta_2, \ \ \tilde{\theta}_2= -((k-1)\theta_1 + k \theta_2).
\ee

The cyclic chamber of the mutated quiver defined by $(\tilde{\theta}_1<0, \tilde{\theta}_2<0)$ is actually 
a small chamber in Fig.\ref{fig:theta} given by
\be 
k \theta_1 + (k+1)\theta_2<0, \ \ (k-1)\theta_1 + k \theta_2>0.
\ee

The partition function will be the PT invariants truncated at certain D2 and D0 charges determined
by $k$.
\be
\mathbb{Z}(u,v) = \prod_{i=1}^{k} (1-(-u)^{i}v^{-1})^{i}
\ee
Therefore if we take large $k$ limit the partition function will approach the PT invariants.
$\spadesuit$

\medskip \medskip

The starting point will be the reversed framed quiver with $\theta_0>0$ and $\theta_1, \theta_2 <0$ as follows:

\begin{equation}
\xymatrix{
& \bullet 0  &  \\
& & \\
\bullet 1 \ar[uur]^{Q_1}  \ar@(ul,dl)[]|{C_1} \ar@/^/@{->>}[rr]^{A_1, A_2} && \bullet 2
\ar@(ur,dr)[]|{C_2} \ar@/^/@{->>}[ll]^{B_1, B_2}  \\
}
\end{equation}

The other advantage to start from this empty chamber is that the Seiberg duality (mutation) can be performed more easily, which
will be explained in the appendix.\footnote{Roughly speaking, when doing the mutation at certain node, we need to look at
the nodes coming into the node on which the mutation is taken. Therefore, making the framing arrow outgoing will simplify
the proof of the tilting property.}
The mutation in such quiver is realized in a very similar way like conifold. Here is a simple illustration of the mutation. A more detailed
treatment on the mutation on affine $A_1$ is given in appendix. We simply quote the result here.

\begin{equation}
\xymatrix{
& \bullet 0 \ar[ddr]^{P_1} & & & & & &  \bullet 0  \ar@{->>}[ddl]_{\tilde{Q}_1, \tilde{Q}_2} &\\
& & \ar[rrr]^{\text{mutation at the first node}} & & & & & &\\
\bullet 1 \ar@(ul,dl)[]|{C_1} \ar@{->>}[uur]^{Q_1,Q_2} \ar@/^/@{->>}[rr]^{A_1, A_2} && \bullet 2  \ar@(ur,dr)[]|{C_2} \ar@/^/@{->>}[ll]^{B_1, B_2}
&  & & & \bullet 1 \ar@(ul,dl)[]|{\tilde{C}_1} \ar@/^/@{->>}[rr]^{\tilde{A}_1, \tilde{A}_2}  & &
\bullet 2
\ar@(ur,dr)[]|{\tilde{C}_2} \ar@/^/@{->>}[ll]^{\tilde{B}_1, \tilde{B}_2}  \ar@/_/[uul] \ar@/_/@<-1ex>[uul] \ar@/_/@<-2ex>[uul]
_{\tilde{P}_1, \tilde{P}_2, \tilde{P}_3}\\
}
\end{equation}

The superpotential for the LHS quiver is given by
\begin{equation}
W= w(C_1)-w(C_2) + C_1 ( B_1 A_1 + B_2 A_2) -C_2 ( A_1 B_1 +A_2 B_2) +B_1 P_1 Q_1 + B_2 P_1 Q_2,
\end{equation}

while the superpotential for the RHS quiver is
\begin{eqnarray}
\tilde{W} = w(\tilde{C}_1) - w(\tilde{C}_2)
&+& \tilde{C}_1 ( \tilde{B}_1 \tilde{A}_1 + \tilde{B}_2 \tilde{A}_2) -\tilde{C}_2 ( \tilde{A}_1 \tilde{B}_1 +\tilde{A}_2 \tilde{B}_2) +
\nonumber \\ &&   \tilde{P}_1 \tilde{A}_1 \tilde{Q}_1+ \tilde{P}_2 (\tilde{A}_2 \tilde{Q}_1 - \tilde{A}_1\tilde{Q}_2)+
\tilde{P}_3 \tilde{A}_2\tilde{Q}_2
\end{eqnarray}

More generically, we have the mutation from $k$ to $k+1$:

\begin{equation}
\xymatrix{
& \bullet 0 \ar[ddr]^{P_1,..,P_k} & & & & & &  \bullet 0  \ar[ddl]_{\tilde{Q}_1,...,\tilde{Q}_{k+1}} &\\
& & \ar[rrr]^{\text{mutation at 1}} & & & & & &\\
\bullet 1 \ar@(ul,dl)[]|{C_1} \ar[uur]^{Q_1,...,Q_{k+1}} \ar@/^/@{->>}[rr]^{A_1, A_2} && \bullet 2  \ar@(ur,dr)[]|{C_2} \ar@/^/@{->>}[ll]^{B_1, B_2}
&  & & & \bullet 1 \ar@(ul,dl)[]|{\tilde{C}_1} \ar@/^/@{->>}[rr]^{\tilde{A}_1, \tilde{A}_2}  & &
\bullet 2
\ar@(ur,dr)[]|{\tilde{C}_2} \ar@/^/@{->>}[ll]^{\tilde{B}_1, \tilde{B}_2}  \ar[uul]_{\tilde{P}_1,..., \tilde{P}_{k+2}}\\
}
\end{equation}

The superpotentials are

\begin{eqnarray}
W &=& w(C_1)-w(C_2) + C_1 ( B_1 A_1 + B_2 A_2) -C_2 ( A_1 B_1 +A_2 B_2) +\nonumber \\
&& Q_1 B_1 P_1 +Q_2 (B_2 P_1-B_1 P_2)+Q_3(B_2 P_2 - B_1 P_3) +\cdots + Q_{k+1} B_2 P_k \\
 \tilde{W} &=& w(\tilde{C}_1) - w(\tilde{C}_2) + \tilde{C}_1 ( \tilde{B}_1 \tilde{A}_1 +
 \tilde{B}_2 \tilde{A}_2) -\tilde{C}_2 ( \tilde{A}_1 \tilde{B}_1 +\tilde{A}_2 \tilde{B}_2) +
\nonumber \\
&&   \tilde{P}_1 \tilde{A}_1 \tilde{Q}_1+ \tilde{P}_2 (\tilde{A}_2 \tilde{Q}_1 - \tilde{A}_1\tilde{Q}_2)+
\tilde{P}_3  (\tilde{A}_2 \tilde{Q}_2 - \tilde{A}_1\tilde{Q}_3) + \cdots + \tilde{P}_{k+2} \tilde{A}_2 \tilde{Q}_{k+1}
\end{eqnarray}

This local obstructed $\mathbb{P}^1$ geometry can be resolved into $n$ isolated
conifold points and all the n $\mathbb{P}^1$s are homologous to each other.
Recall that  the authors in \cite{AK} choose the same basis for this geometry as 
the conifold in the Ext group computation. Under the mutation the $\theta$ parameters
should also change according to (\ref{eq:FI}). Moreover, one can perform a similar 
wallcrossing analysis as \cite{JM} and should obtain the same chamber structure. 

With all the pieces of information, our educational expectation would be 
that in every chamber the BPS states partition function should be n-th power of that of 
the conifold. 
In the next section we will classify all the fixed points in the moduli space.

\section{Classification of the Torus Fixed Points}
\label{sec:class}
In this section we classify all the torus fixed points in the moduli space.
As we will see later the ERC is $n$ copies of conifold ERC of length $k$.
We denote this ERC by \textit{"$(n,k)$ (in)finite ERC"} for short.
We find that  the fixed points are classified by filtered pyramid partitions of $(n,k)$ (in)finite ERC.
At first glance, it seems strange because we really want $n$ independent copies of the
conifold answer. If the fixed points are classified by filtered partitions and the partitions are
correlated, how can we reproduce the desirable answer?

But we have to be more careful here because the torus actions in the geometry only fix invariant the F-term relations, not
the superpotential itself, which renders the deformation complex not symmetric and self-dual.
Because of this the local contribution is actually not simply $(-1)^{\text{dim} T_f \mathcal{M}}$
but rather a rational function of the localization parameters. In the end we will show by examples
that the integrability reappear after grouping together all the fixed points with same dimension vectors.

\medskip

Consider the torus action on moduli space $\mathcal{M}_{\textbf{v}}$
of framed, cyclic representations of $\mathcal{A}$ with a single framing $m$,
where $\vec{v}=(v_1,v_2)$ is the dimension vector:
\begin{equation}
\begin{split}
(t_1,t_2)\in (C^*)^2: \mathcal{A}& \rightarrow \mathcal{A}\\
(A_1,A_2,B_1,B_2,C_1,C_2)& \mapsto (t_1^{-1} t_2^n A_1, t_1 t_2^n A_2, t_1 B_1, t_1^{-1} B_2, t_2 C_1, t_2 C_2)\end{split}
\end{equation}

\begin{figure}[ht]
\centering
\epsfig{file=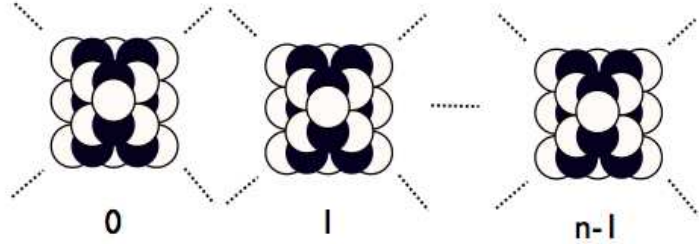, width=10cm}
\caption{This is the ERC for affine $A_1$ quiver with superpotential degree (n+1)
and 1 framing arrows, which is not reversed. It is denoted by $(n,1)$ infinite ERC.}
\label{fig:ERCinfinite}
\end{figure}

\textbf{Proposition} \textit{The fixed points under torus actions
above are isolated and classified by filtration of pyramid
partitions of $(n, k=1)$ infinite ERC.} The $(n,k=1)$ infinite ERC is shown in 
Fig.\ref{fig:ERCinfinite}.

Proof: Denote the Klebanov-Witten quiver path algebra $\mathcal{A}_c$ which
only involves $A_i$ and $B_i$ and has the same relation (\ref{eq:ABA}).
Note that the algebra $\mathcal{A}$ is $\mathbb{Z}_n$-graded by the power of
$C_i$,\footnote{It is not graded strictly speaking because $C_i^n$
is not identity.} one can always move $C_i$ to the very left by
using (\ref{eq:commuteC}).

\be
\mathcal{A}=\bigoplus_{k=0}^{n-1}\mathcal{A}_k=\bigoplus_{k=0}^{n-1}(C_1+C_2)^k\mathcal{A}_c
\ee

There is a natural projection $\pi_k:\mathcal{A}_k\rightarrow \mathcal{A}_c$. So is the
framed cyclic module $M$: \be{M=\bigoplus_{k=0}^{n-1}M_k}\ee where
$M_k=(C_1+C_2)^kM_c$ for some framed $\mathcal{A}_c$-module $M_c$ with the
same framing $m$.

By a similar argument as in \cite{Szendroi}, one can show that the
fixed point is in 1-1 correspondence with the $m$-annihilator ideal
of $Q$ generated by monomial: 
\be 
\mathcal{A} \supset I=\bigoplus_{k=0}^{n-1} I_k
\ee

If $\alpha\in I_k$, then $(C_1+C_2)\alpha\in I_{k+1}$. Thus we
conclude: \be\pi_1I_1\subset \pi_2I_2\subset\cdots\subset
\pi_kI_k\subset\cdots\ee

Moreover each $I_k$ is classified by a pyramid partition consisting
of black and white stones denoting one-dimensional subspaces of
given toric weights \cite{Szendroi}. Roughly speaking $I_k$
tells how to truncate an infinite pyramid partition from below,
therefore we conclude that the fixed point corresponds to a nested
pyramid partition of length $n$, where the sum of all white(black)
stones is $v_1$($v_2$).

For mutated quiver, we have similar torus
action:

\be(A_1,A_2,B_1,B_2,C_1,C_2,P_1,P_2,\dots, P_k, Q_1,Q_2, \dots,
Q_{k+1})\ee having weights \be{ (  \underbrace{ t_1^{-1}t_2^n,t_1
t_2^n }_{A_1, A_2}, \underbrace{t_1,t_1^{-1}}_{B_1,
B_2},\underbrace{t_2,t_2}_{C_1, C_2},
\underbrace{1,t_1^2,t_1^4,\dots}_{P_1,...,P_k}, \underbrace{t_1 t_2,
t_1^{-1} t_2, t_1^{-3} t_2 ,\dots}_{Q_1,...,Q_{k+1}})}\ee

\begin{figure}[ht]
\centering
\epsfig{file=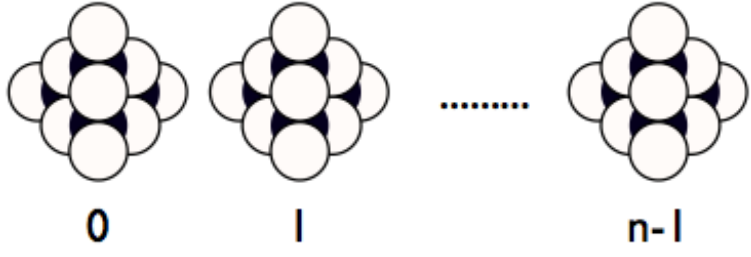, width=10cm}
\caption{This is the ERC for affine $A_1$ quiver with superpotential degree (n+1)
and 3 framing arrows, denoted by $(n,3)$ finite ERC.}
\label{fig:ERCfinite}
\end{figure}

\begin{figure}[ht]
\centering
\epsfig{file=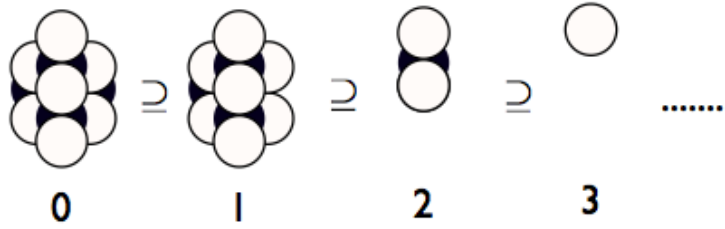, width=10cm}
\caption{An example of the filtered pyramid partitions of $(n,3)$ finite ERC. ($n \geq 3$) }
\label{fig:filtered}
\end{figure}

The algebra is also graded, but this time we have to impose
cyclicity first to kill the $Q_i$'s, then one can similarly project
to subalgebra of mutated conifold quiver, whose torus fixed points
are classified by finite pyramid partition. Thus we conclude:

\textbf{Proposition} \textit{Mutated quiver at step $k$ has a torus
action whose fixed points are classified by filtration of pyramid
partitions of $(n, k)$ (in)finite ERC.} Fig.\ref{fig:ERCfinite} shows the $(n,3)$ finite ERC.
Fig.\ref{fig:filtered} represents a fixed point in $(n,3)$ finite ERC.

\medskip

Now we provide some examples with different dimension vectors $\vec{v}$ and general value of $n$.
The basis of the quiver with $k$ framing arrows
is ${ \mathcal{O}_C(-k-1)[1], \mathcal{O}_C(-k-2) }$. Translating it into brane charges we have
$\{ D2 + (k+1) \ \bar{D0} , \bar{D2} + k \ D0 \}$.
We then denote $\Gamma = (\bar{D2} \ \text{charge}, D0 \ \text{charge} ).$

{\bf Example 1}  \ \  $k=2$, general $n$, $\vec{v}=(1,0)$,  $\Gamma =(1,2)$. Two fixed points are
$p_1 = \{V_1=1, V_2=0\}$ and $p_2 = \{V_1=t_1^{-2},V_2=0\}$.
The deformation complex (\ref{eq:complex1})(\ref{eq:complex2}) simplifies to
\begin{eqnarray}
0 \rightarrow && \text{Hom}(V_1,V_1) \rightarrow \text{Hom}( V_1, V_1 \otimes t_2) \oplus \text{Hom}(\mathbb{C}, V_1)
\oplus \text{Hom}(\mathbb{C}, V_1 \otimes t_1^2) \rightarrow \nonumber \\ &&
\text{Hom}( V_1, V_1 \otimes t_2^n) \oplus \text{Hom}(V_1, \mathbb{C} \otimes t_2^{n+1})
\oplus \text{Hom}(V_1, \mathbb{C} \otimes t_1^{-2} T_2^{n+1} ) \rightarrow \nonumber \\
&&\text{Hom}(V_1,V_1 \otimes t_2^{n+1})  \rightarrow 0
\end{eqnarray}

Abusing the notation, we first compute
\be
\mathcal{T} = \mathcal{T}_0 -\mathcal{T}_1 +\mathcal{T}_2 -\mathcal{T}_3 = V_1^{*} \otimes V_1 (1-t_2+t_2^{n}-t_2^{n+1}) -V_1(1+t_1^2) + V_1^{*}(1+t_1^{-2})t_2^{n+1}.
\ee
for each of the fixed point and then perform the substitution
\be
\mathcal{T} = \sum k_{(n_1,n_2)} t_1^{n_1} t_2^{n_2} \to P(\alpha_1, \alpha_2)=\prod (n_1 \alpha_1 + n_2 \alpha_2)^{k_{(n_1,n_2)}}.
\ee

In this case
\begin{eqnarray}
P_{p_1} = -n + \frac{n+1}{2} \frac{\alpha_1}{\alpha_2}\\
P_{p_2} = -n - \frac{n+1}{2} \frac{\alpha_1}{\alpha_2}\\.
\end{eqnarray}
Therefore, we obtain the Euler character of the moduli space
\be
e(\mathcal{M}_{(1,2)}) =-2n.
\ee
Here we specify the moduli by the Dbrane charges for later convenience.

\begin{figure}[ht]
\centering
\epsfig{file=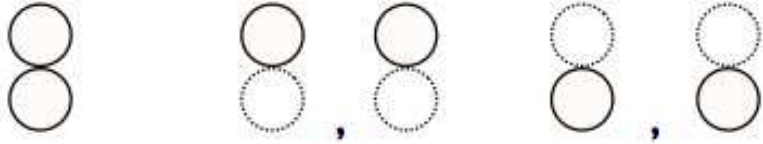, width=10cm}
\caption{This figure shows the three fixed point in Example 2. }
\label{fig:fixed2}
\end{figure}

{\bf Example 2} \ \ Fig. \ref{fig:fixed2}. $k=2$, general $n$, $\vec{v}=(2,0)$, $\Gamma =(2,4)$.
We find three fixed points: $p_1 = \{V_1=1+t_1^{-2}, V_2=0 \}$, $p_2 = \{ V_1=1+t_2^{-1}, V_2=0 \}$
and $p_3 = \{ V_1=t_1^{-2}+t_1^{-2} t_2^{-1}, V_2=0 \}$
The final result is
\be
e(\mathcal{M}_{(2,4)}) = -n + 2n^2 .
\ee

\begin{figure}[ht]
\centering
\epsfig{file=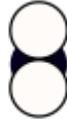, width=0.8cm}
\caption{The single fixed point for dimension vector $\vec{v}=(2,1)$.}
\label{fig:fixed3}
\end{figure}

{\bf Example 3} \ \ Fig. \ref{fig:fixed3}. $k=2$, general $n$, $\vec{v}=(2,1)$, $\Gamma =(1,1)$.
The single fixed in this case is $p_1 = \{V_1=1+t_1^{-2}, V_2=t_1^{-1} \}$. This is the first
nontrivial example in which $V_2 \neq 0$. We have to use the whole complex (\ref{eq:complex1})
 (\ref{eq:complex2}) and obtain,
\be
e(\mathcal{M}_{(1,1)}) = n .
\ee

\begin{figure}[ht]
\centering
\epsfig{file=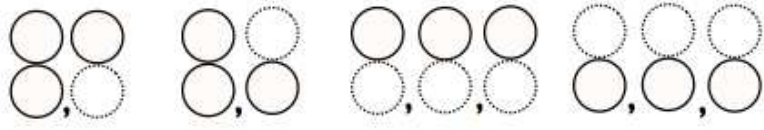, width=10cm}
\caption{4 fixed points for $\vec{v}=(3,0)$.}
\label{fig:fixed4}
\end{figure}

{\bf Example 4} \ \ Fig. \ref{fig:fixed4}. $k=2$, general $n$, $\vec{v}=(3,0)$, $\Gamma =(3,6)$
$p_1 = \{V_1=1+t_1^{-2}+t_2^{-1}, V_2=0 \}$, $p_2 = \{V_1=1+t_1^{-2}+t_1^{-2}t_2^{-1}, V_2=0 \}$,
$p_3 = \{V_1=1+t_2^{-1}+t_2^{-2}, V_2=0 \}$, $p_3 = \{V_1=(1+t_2^{-1}+t_2^{-2})t_1^{-2}, V_2=0 \}$.
\be
e(\mathcal{M}_{(3,6)}) = -\frac{2n(2n-1)(2n-2)}{6} .
\ee

{\bf Compare with the expected result.}
Now let us compare these examples with the expected answers. 
Partition function for cyclic representations of that mutated conifold quiver with $k$ framing arrows
is\cite{CJ}

\be
\mathcal{Z}^{\text{conifold}}_k=\sum_{p,q}D(p,q) u^{q}v ^{p}=\prod_{j=1}^k (1-(-u)^jv)^j
\ee

where
$D(p,q)$ is the virtual Euler character of quiver moduli space of
$\bar{D2}/D0$ charge $(p,q)$. In affine $A_1$ quiver with superpotential deformation $w(C_i)$
of degree $n+1$, we expect to get
\be
\mathcal{Z}^{w}_k=(Z^{\text{conifold}}_k)^n=\prod_{j=1}^k(1-(-u)^j v )^{jn}.
\ee

In the explicit examples presented above, we all have $k=2$. Therefore we should compare them
with the following:
\begin{eqnarray}
\mathcal{Z}_{k=2} = && (1-(-u)v)^{n}(1+u^2v)^{2n} = (1+n uv + \binom{n}{2} u^2 v^2 + \binom{n}{3} u^3v^3 + \cdots) \nonumber \\
&& (1- 2n u^2v + \binom{2n}{2} u^4 v^2 - \binom{2n}{3} u^6v^3 + \cdots)
\end{eqnarray}

The Euler characters of the examples can be read off as follows:
\be
e(\mathcal{M}_{(1,2)}) =-2n, \ \ e(\mathcal{M}_{(2,4)}) = -n + 2n^2, \ \ e(\mathcal{M}_{(1,1)}) = n, \ \ e(\mathcal{M}_{(3,6)}) = -\frac{2n(2n-1)(2n-2)}{6} .
\ee
These computations provide strong favorable evidences for our classification of the fixed points.

\section{A New Type of Melting Crystal Model?}
\label{sec:melt}
The statistical model of the melting crystal model was first proposed as a realization of quantum
gravitational path integral over the fluctuations of the K\"ahler geometries \cite{qfoam}.

Recently there are again many activities in constructing the statistical model of melting crystal
to count the BPS state bound states in noncompact Calabi-Yau manifolds. The smooth classical
emerges as a thermodynamic limit of the statistical mechanical model of crystal melting \cite{OY}.
Mathematically speaking, each atoms or melting crystal configuration corresponds to a
torus fixed point in the moduli space of the quiver representation. Since the geometry is
toric, every fixed point should contribute either $+1$ or $-1$, depending on the dimension
vector of the quiver theory.

In our example, we find the weights of the fixed point are rational functions, rather than
integers. This suggests there should exist a large class of melting crystal model with more
general weights. There exist many other obstructed local $\mathbb{P}^1$ geometries,
which are not toric and have two natural $(\mathbb{C}^{*})^2$ actions.
For example, we can construct obstructed local $\mathbb{P}^1$ with normal bundle
$\mathcal{O}(1) \oplus \mathcal{O}(-3)$ by gluing two patches of $\mathbb{C}^3$,
with coordinates $(x,y_1,y_2)$ and $(w,z_1, z_2)$ by the following transition functions\cite{Zhou}:

\begin{equation}
z_1 = x^3y_1+y_2^2+ x^2 y_2^{2n+1}, \ \ \ z_2 = y_2/x, \ \ \ w=1/x.
\end{equation}

It would be also interesting to study the BPS state counting in this class of geometries and say something
useful about the weight assignment in the study of quantum geometries.

\section{Conclusion}
\label{sec:conclusion}

In this paper we study the BPS state counting in the geometry of local obstructed
curve with normal bundle $\mathcal{O} \oplus \mathcal{O}(-2)$.
We find that the D6-D2-D0 bound states have a framed affine $A_1$ quiver description
with degree $(n+1)$ superpotential deformation.

We then develop a new way to mutate this framed quiver with adjoint fields and obtain the
new superpotential and the new framing structure after the mutation. The mutation
on the quiver with both adjoint fields and framing is nontrivial and is presented 
in Appendix ABCD. Using this quiver description along with the mutation, 
we can bring the original framed quiver
in any chambers to a certain mutated quiver in the cyclic chamber.

The new superpotential and cyclicity in the mutated quiver will simplify dramatically the
classification of the fixed points. We find that the fixed points are classified
by \textit{filtered pyramid partitions of $(n,k)$ (in)finite ERC.}
We verify this classification by doing explicit computations of the local contributions of the fixed
points. In this nontoric geometry, the deformation complex will cease to be selfdual and therefore
the integrability only appears after grouping together the fixed points with the same dimension
vector.

This method of mutating back to the cyclic chamber
provides a well-defined method for computing generalized Donaldson-Thomas invariant in different chambers.
This general idea can be generalized to other affine ADE quiver theories. We present some affine $A_n$ results
in the appendix \ref{append:D} and leave the DE-type cases for future study. 
Finally we comment on the implications on the melting crystal models and mention possible future
directions.


\medskip

{\bf Acknowledgments:} We would like to thank Emanuel Diaconescu
for suggesting the problem and for his patient guidance throughout the project.
WYC and GP are supported by DOE grant DE-FG02-96ER40959.

\newpage

\appendix
In the appendix we first briefly review the procedure for
performing the mutation to the quiver theory. After that we give explicit
examples of the mutation for affine A-type quiver without and with framing.
We find that the mutation of the affine $A_1$ quiver with superpotential deformations
works pretty much the same way as the conifold \cite{CJ}. The mutation of 
affine $A_n$ quiver is given in the subsequent section. We then give the deformation
complex for the affine $A_1$ case to compute the local contribution of the torus fixed points.

\section{Seiberg Duality and Mutation}
\label{append:A}

The section is a crash review on the mutation, following \cite{vitoria}.
The Seiberg duality or mutation on quiver theory is a tilting procedure and therefore
is an equivalence of the derived categories. In order to check if a complex is
tilting one needs to compute the morphisms in the homotopy category between
the objects, namely, the complex formed by the projective objects.

First consider a quiver with superpotential and let $\mathcal{A}$ be the path algebra
of the quiver and $\mathcal{Q}$ the abelian category of $\mathcal{A}$-module.
Let $K^{b}( \mathcal{Q} )$ be the homotopy category, where the morphisms
are the homotopy classes of the chain maps.
A tilting complex over $\mathcal{A}$ is an object $T$ in $K^{b}(\mathcal{Q})$, satisfying
the following two properties:
\begin{itemize}
\item[$(i)$] $Hom_{ K^{b}(\mathcal{Q}) }(T, T[n])=0$ for all $n \neq 0$.
\item[$(ii)$] $T$ generates $K^{b}(\mathcal{Q})$ as a triangulated category.
\end{itemize}

Our procedure for performing the mutation on framed affine $A_n$ quiver theory at
k-th vertex, where the framing arrows are outgoing,
is to replace the one term complex at k-th vertex
\begin{equation}
0 \rightarrow \underline{\mathcal{P}_k} \rightarrow 0 \nonumber
\end{equation}
by
\begin{equation}
0 \rightarrow \underline{\oplus_{j \to k} \mathcal{P}_j} \rightarrow \mathcal{P}_k \rightarrow 0 \nonumber
\end{equation}

where the underline means the zeroth position of the complex and
$\mathcal{P}_k$ is the projective module consisting of all the paths in the path algebra ending
on $k$-th node. It is easy to check that it is projective.

In this case we define the tilting complex

\begin{equation}
T= \oplus_{i=1}^{n} T_i \nonumber
\end{equation}
where
\begin{eqnarray}
T_i : \ \ 0 \rightarrow \underline{\mathcal{P}_i} \rightarrow 0, \ \ i \neq k \nonumber \\
T_k : \ \ 0 \rightarrow \underline{\oplus_{j \to k} \mathcal{P}_j} \rightarrow \mathcal{P}_k \rightarrow 0
\end{eqnarray}

Most of the morphisms in $Hom_{K^{b}( \mathcal{Q} )}(T, T[n])$ are trivial up to the homotopy except the following
two diagrams, which potentially might represent certain nontrivial morphisms in $Hom_{K^{b}( \mathcal{Q} )}(T, T[n])$.

\be
\label{eq:prob1}
\xymatrix{
0 \ar[r] & \oplus_{j \to k} \mathcal{P}_j \ar[r] \ar[d] & \mathcal{P}_k \ar[r] \ar[d] & 0 \\
& 0 \ar[r]  & \mathcal{P}_i \ar[r] & 0\\}
\ee

\be
\label{eq:prob2}
\xymatrix{
0 \ar[r] & \oplus_{j \to k} \mathcal{P}_j \ar[r] \ar[d] & \mathcal{P}_k \ar[r] \ar[d] & 0 &\\
& 0 \ar[r]  & \oplus_{j \to k} \mathcal{P}_j  \ar[r] & \mathcal{P}_k \ar[r] & 0 \\}
\ee

(\ref{eq:prob1}) is used to compute $Hom(T_k, T_i[-1])$, while
(\ref{eq:prob2}) is the element in  $Hom(T_k, T_k[-1])$.

A morphism in $Hom(T_k, T_i[-1])$ means a mapping $\alpha : \mathcal{P}_k \to \mathcal{P}_i$ such that
for every mapping $\beta : \oplus_{j \to k} \mathcal{P}_j   \to \mathcal{P}_k$
we have $\beta \circ \alpha =0$, do to the commutativity of
the diagram (\ref{eq:prob1}). This condition is very restricted and we can not find such morphisms
in affine $A_n$ quiver, if we perform the mutation at the node with outgoing framing arrows.
The second tilting property can be easily verified too.

\section{A-type Quiver without Framing}
\label{append:B}
Although Seiberg duality with adjoint fields has been studied in
\cite{Kutasov} \cite{BD}, their procedures are not exactly applicable for our purpose.
We find that we have to resort to the reflection operation in \cite{CH} \cite{vafa1}
and make suitable adjustment due to the presence of the new framing arrows.
In this section we will present the reflection operation in terms of the
aforementioned tilting complex.
We consider the affine $A_2$ quiver without framing:

\begin{equation}
\xymatrix{
& \bullet 1 \ar@(ul,ur)[]|{C_1} \ar@/^/[ddl] \ar@/^/[ddr]  &   \\
& & \\
\bullet 2 \ar@(ul,dl)[]|{C_2} \ar@/^/[rr] \ar@/^/[uur] && \bullet 3 \ar@(ur,dr)[]|{C_3} \ar@/^/[ll] \ar@/^/[uul]   \\
}
\end{equation}

$X_{ij}$ denotes the mapping from $j$-th node to $i$-th node.
And $C_i$ is the adjoint field at $i$-th node. The relations are given by \cite{vafa1}
\begin{eqnarray}
&& X_{i,i-1} X_{i-1,i} - X_{i,i+1} X_{i+1,i} =  w_i(C_i) \\
&& X_{ij} C_j = C_i X_{ij} \\
&& \sum_{i} d_i w_i =0, \ \ \text{$d_i$:Dynkin index at i-th node.}
\end{eqnarray}

Let us mutate the 2nd vertex, which amounts to replacing
\be
\xymatrix{
0 \ar[r] & \mathcal{P}_2 \ar[r] & 0 }
\ee
by
\be
\xymatrix{
0 \ar[r] & \mathcal{P}_1 \oplus \mathcal{P}_3 \ar[r]^{(X_{21},X_{23})}  & \mathcal{P}_2 \ar[r]  & 0 }.
\ee

It has been shown in \cite{CH} \cite{vafa1} that the mutation in this quiver
is actually a Weyl reflection with respect to the mutated node.
For example, after mutation we should have
\be
\tilde{X}_{21} \tilde{X}_{12} - \tilde{X}_{23} \tilde{X}_{32} = - w_2(C_2)
\ee
where the $\tilde{X}$ fields are the ones after mutation.

Since we are simply interested in the consistency between mutation procedure
proposed here and the results in \cite{CH} \cite{vafa1}, we present
the morphism after the mutation explicit.

The following diagram give $\tilde{X}_{21}$ and $\tilde{X}_{12}$:
\be
\xymatrix{
0 \ar[r] & \mathcal{P}_1 \oplus \mathcal{P}_3 \ar[r]^{(X_{21},X_{23})} \ar[d]^{(id,0)} & \mathcal{P}_2 \ar[r] \ar[d] & 0 & \\
0 \ar[r] & \mathcal{P}_1 \ar[r] \ar[dd]_{ \left( \begin{array}{c}
-w_2 + X_{12}X_{21}\\X_{32}X_{21}
 \end{array} \right) } & 0 \ar[dd] & & \\
& & & & \\
0 \ar[r] & \mathcal{P}_1 \oplus \mathcal{P}_3 \ar[r]^{(X_{21},X_{23})} & \mathcal{P}_2 \ar[r] & 0 &
}
\ee
\\ \\

Here are $\tilde{X}_{23}$ and $\tilde{X}_{32}$.
\be
\xymatrix{
0 \ar[r] & \mathcal{P}_1 \oplus \mathcal{P}_3 \ar[r]^{(X_{21},x_{23})} \ar[d]^{(0,id)} & \mathcal{P}_2 \ar[r] \ar[d] & 0 & \\
0 \ar[r] & \mathcal{P}_3 \ar[r] \ar[dd]_{ \left( \begin{array}{c}
X_{12}X_{23}\\ -w_2 + X_{32}X_{23}
 \end{array} \right) } & 0 \ar[dd] & & \\
& & & & \\
0 \ar[r] & \mathcal{P}_1 \oplus \mathcal{P}_3 \ar[r]^{(X_{21},X_{23})} & \mathcal{P}_2 \ar[r] & 0 &
}
\ee

\begin{table}[ht]
\centering 
\begin{tabular}{c c c c} 
\hline\hline 
 & $T_1$ & $T_2$ & $T_3$ \\ [0.5ex] 
\hline 
$T_1$ & id &  $\left( \begin{array}{c}
-w_2(C_2) + X_{12}X_{21}\\X_{32}X_{21}
 \end{array} \right)$ & $X_{31}$\\
$T_2$ &  $(id,0)$ & $\left( \begin{array}{cc}
id & 0\\
0 & id  \\
\end{array} \right)$   & $(0,id)$ \\
$T_3$ & $X_{13}$ & $\left( \begin{array}{c}
X_{12}X_{23}\\ -w_2(C_2) + X_{32}X_{23}
 \end{array} \right)$  & id\\ [1ex]
\hline
\end{tabular}
\caption{Table for all the bifundamental field after the mutation.}
\label{table:morphism}
\end{table}

Now the reflecting procedure can be checked very straightforwardly. For example,
\be
\tilde{X}_{21} \tilde{X}_{12} - \tilde{X}_{23} \tilde{X}_{32} = - w_2(C_2)
\ee
is indeed satisfied after new fields are plugged in.

\section{Affine $A_1$ Quiver with Framing}
\label{append:C}

This framed quiver in this section will be an affine $A_1$ quiver with a framing arrow connecting
the first vertex in the quiver. The mutation of this quiver is a generalization of \cite{CH}.
Our procedure is that we apply the mutation to the node with the outgoing framing arrows only.
We explain the procedure by the following example.

\begin{equation}
\label{eq:quiver1}
\xymatrix{
& \bullet 0 \ar[ddr]^{P_1} &  \\
& & \\
\bullet 1 \ar@(ul,dl)[]|{C_1} \ar@{->>}[uur]^{Q_1,Q_2}
\ar@/^/@{->>}[rr]^{A_1, A_2} && \bullet 2  \ar@(ur,dr)[]|{C_2} \ar@/^/@{->>}[ll]^{B_1, B_2}  \\
}
\end{equation}

\begin{equation}
W= w(C_1)-w(C_2) + C_1 ( B_1 A_1 + B_2 A_2) -C_2 ( A_1 B_1 +A_2 B_2) +B_1 P_1 Q_1 + B_2 P_1 Q_2
\end{equation}
The relations coming from various fields are
\begin{eqnarray}
\label{eq:oldrel}
&& C_1 B_i = B_i C_2 \nonumber \\
&& A_2 C_1 - C_2 A_2 + P_1 Q_2 =0 \nonumber \\
&& A_1 C_1 - C_2 A_1 + P_1 Q_1 =0 \nonumber \\
&& w'(C_1) + B_1 A_1 + B_2 A_2 = 0 \nonumber \\
&& w'(C_2) + A_1 B_1 + A_2 B_2 = 0 \nonumber \\
&& Q_1 B_1 + Q_2 B_2 =0, \ B_1 P_1 = B_2 P_1= 0 \nonumber \\
\end{eqnarray}

The complex for the first node now is being replaced by the two term complex,
\begin{equation}
0 \to \underline{ \mathcal{P}_2 + \mathcal{P}_2 } \to \mathcal{P}_1 \to 0
\end{equation}

Here we will adapt a very \textit{straightforward} approach. Namely
we will present all the new fields in terms of old fields and write down the
new \textit{would-be} quiver with new \textit{would-be} superpotential.
And then by direct computation we can see that the new relations implied
by the new \textit{would-be} superpotential are satisfied, which concludes
our mutation procedure.

First the new $\tilde{A}_1$ and $\tilde{A}_2$ are given by

\begin{equation}
\xymatrix{
0 \ar[r] & \mathcal{P}_2 +  \mathcal{P}_2 \ar[d]^{(1,0),(0,1)} \ar[r]^{(B_1,B_2)} & \mathcal{P}_1 \ar[r] \ar[d]^{0}  & 0 & \\
0 \ar[r] & \mathcal{P}_2 \ar[r] &  0 & &
}
\end{equation}

We can as well denote $\tilde{A}_1 = ((1,0),0)$ and $\tilde{A}_2=((0,1),0)$
Similarly $\tilde{B}_1$ and $\tilde{B}_2$ are the chain maps of the following
diagrams. One can easily check that the diagram commutes after using the
relations $A_1 B_i A_2 = A_2 B_i A_1$ and $B_1 A_i B_2 = B_2 A_i B_1$.

\begin{displaymath}
\xymatrix{
0 \ar[r] & \mathcal{P}_2 \ar[d] \ar[r] &  0 \ar[d] & & \\
0 \ar[r] & \mathcal{P}_2 +  \mathcal{P}_2 \ar[r]^{(B_1,B_2)} & \mathcal{P}_1 \ar[r]  & 0 & \\
}
\end{displaymath}

\begin{equation}
\tilde{B}_1 = ( \left( \begin{array}{c}
A_2 B_2\\ -A_2 B_1
 \end{array} \right),0)\ \ , \ \ \   \tilde{B}_2 = ( \left( \begin{array}{c}
-A_1 B_2\\ A_1 B_1
 \end{array} \right),0)
\end{equation}

The new adjoint field $\tilde{C}_1$ becomes:

\begin{displaymath}
\xymatrix{
0 \ar[r] & \mathcal{P}_2 + \mathcal{P}_2 \ar[r] \ar[dd]_{\left( \begin{array}{cc}
C_2 & 0 \\ 0 & C_2
 \end{array} \right)} & P_1 \ar[r] \ar[dd]^{C_1} & 0 & \\
& & & & \\
0 \ar[r] & \mathcal{P}_2 + \mathcal{P}_2 \ar[r] & \mathcal{P}_1 \ar[r] & 0 &
}
\end{displaymath}

\begin{equation}
\tilde{C}_1 = ( \left( \begin{array}{cc}
C_2 & 0 \\ 0 & C_2
 \end{array} \right),C_1)
\end{equation}

The maps $\tilde{Q}_1$ and $\tilde{Q}_2$ are chosen in following way.
\begin{displaymath}
\xymatrix{
0 \ar[r] & \mathcal{P}_0 \ar[d] \ar[r] &  0 \ar[d] & & \\
0 \ar[r] & \mathcal{P}_2 +  \mathcal{P}_2 \ar[r]^{(B_1,B_2)} & \mathcal{P}_1 \ar[r]  & 0 & \\
}
\end{displaymath}

\begin{equation}
\tilde{Q}_1 = ( \left( \begin{array}{c}
P_1\\ 0
 \end{array} \right),0)\ \ , \ \ \   \tilde{Q}_2 = ( \left( \begin{array}{c}
0\\ P_1
 \end{array} \right),0)
\end{equation}

The commutativity of the diagram is assured by the relation $B_1 P_1 = B_2 P_1 =0$.

The new \textit{would-be} quiver can now be summarized in this quiver diagram.
\begin{equation}
\xymatrix{
& \bullet 0 \ar@/_/[ddr]_{P_1} \ar@{->>}[ddl]_{\tilde{Q}_1, \tilde{Q}_2} &  \\
& & \\
\bullet 1 \ar@(ul,dl)[]|{\tilde{C}_1} \ar@/^/@{->>}[rr]^{\tilde{A}_1, \tilde{A}_2} && \bullet 2
\ar@(ur,dr)[]|{\tilde{C}_2} \ar@/^/@{->>}[ll]^{\tilde{B}_1, \tilde{B}_2}  \ar@/_/[uul] \ar@/_/@<-1ex>[uul] \ar@/_/@<-2ex>[uul] \ar@/_/@<-3ex>[uul]
_{[Q_iB_j]}\\
}
\end{equation}

where $[...]$ means that we should treat it as a single field in the new quiver. They are
simply mesonic fields in the Seiberg duality. Note that $C_2= \tilde{C}_2$.
Now we give the \textit{would-be} superpotential in terms of the new field after the mutation
and then check that the new relations are satisfied, given that the relations coming
from the old superpotential hold.

Here is the new superpotential:
\begin{eqnarray}
\tilde{W} &=& w(\tilde{C}_1) - w(\tilde{C}_2)
+ \tilde{C}_1 ( \tilde{B}_1 \tilde{A}_1 + \tilde{B}_2 \tilde{A}_2) - \tilde{C}_2 ( \tilde{A}_1 \tilde{B}_1 +\tilde{A}_2 \tilde{B}_2) +
[Q_1 B_1] P_1 + [Q_2 B_2] P_1 \nonumber \\ &&  [Q_1 B_1] \tilde{A}_1 \tilde{Q}_1 +  [Q_1 B_2]\tilde{A}_2 \tilde{Q}_1+
[Q_2 B_1]\tilde{A}_1\tilde{Q}_2+ [Q_2 B_2]\tilde{A}_2\tilde{Q}_2
\end{eqnarray}

Note that the field $P_1$ can be integrated out and we have $[Q_1 B_1] + [Q_2 B_2] =0$. We can substitute
this relation back and obtain,
\begin{eqnarray}
\tilde{W} &=& w(\tilde{C}_1) - w(\tilde{C}_2)
+ \tilde{C}_1 ( \tilde{B}_1 \tilde{A}_1 + \tilde{B}_2 \tilde{A}_2) - \tilde{C}_2 ( \tilde{A}_1 \tilde{B}_1 +\tilde{A}_2 \tilde{B}_2) +
\nonumber \\ &&  [Q_1 B_1] \tilde{A}_1 \tilde{Q}_1 +  [Q_1 B_2]\tilde{A}_2 \tilde{Q}_1+
[Q_2 B_1]\tilde{A}_1\tilde{Q}_2- [Q_1 B_1]\tilde{A}_2\tilde{Q}_2
\end{eqnarray}

\medskip

Now let us list all the relations for $\tilde{A}_i$, $\tilde{B}_i$, $\tilde{C}_1$, $\tilde{C}_2$, $\tilde{Q}_i$ and mesonic fields
$[Q_iB_j]$.

\begin{eqnarray}
\tilde{A}_i: \ \ \ &&\tilde{C}_1 \tilde{B}_i - \tilde{B}_i \tilde{C}_2 + \tilde{Q}_1 [Q_1 B_i] +\tilde{Q}_2 [Q_2 B_i]=0 \nonumber \\
\tilde{B}_i: \ \ \ &&\tilde{A}_i \tilde{C}_1 - \tilde{C}_2 \tilde{A}_i = 0 \nonumber \\
\tilde{C}_1: \ \ \ && w'(\tilde{C}_1) + \tilde{B}_1 \tilde{A}_1 + \tilde{B}_2 \tilde{A}_2 =0  \nonumber \\
C_2: \ \ \ && w'(\tilde{C}_2) + \tilde{A}_1 \tilde{B}_1 +\tilde{A}_2 \tilde{B}_2 =0 \nonumber \\
\tilde{Q}_i: \ \ \ &&  [Q_i B_1] \tilde{A}_1 +  [Q_i B_2] \tilde{A}_2=0 \nonumber \\
\lbrack Q_i B_j \rbrack && \tilde{A}_1 \tilde{Q}_2 = \tilde{A}_2 \tilde{Q}_1 =0 , \ \ \tilde{A}_1 \tilde{Q}_1 -\tilde{A}_2 \tilde{Q}_2=0  
\end{eqnarray}

The relations for $\tilde{B}_i$, $\tilde{C}_2$, and $[Q_iB_j]$ are pretty straightforward to
check. For example, $\tilde{A}_1 \tilde{Q}_1 -\tilde{A}_2 \tilde{Q}_2 = (1,0) \binom{P_1}{0} - (0,1) \binom{0}{P_1}= 0$.
Now let us look at the relations for $\tilde{C}_1$ and $\tilde{Q}_i$.

\begin{eqnarray}
w'(\tilde{C}_1) + \tilde{B}_1 \tilde{A}_1 + \tilde{B}_2 \tilde{A}_2 &=& (\left( \begin{array}{cc}
w'(C_2) & 0 \\ 0 & w'(C_2)
 \end{array} \right),w'(C_1)) + (\left( \begin{array}{cc}
A_2 B_2 & -A_1 B_2 \\ -A_2 B_1  & A_1 B_1
 \end{array} \right),0) \nonumber \\
&&( \left( \begin{array}{cc}
-A_1 B_1 & -A_1 B_2 \\ -A_2 B_1  & -A_2 B_2
 \end{array} \right), w'(C_1))
\end{eqnarray}

This chain map is in fact homotopic to zero because of the following diagram.

\begin{equation}
\xymatrix{
0 \ar[r] & \mathcal{P}_2 + \mathcal{P}_2 \ar[r]^{(B_1,B_2)} \ar[dd]_{\left( \begin{array}{cc}
-A_1 B_1 & -A_1 B_2 \\ -A_2 B_1  & -A_2 B_2
 \end{array} \right)} & \mathcal{P}_1 \ar[r] \ar[dd]^{w'(C_1)} \ar[ddl]|{\binom{-A_1}{-A_2}}& 0 & \\
& & & & \\
0 \ar[r] & \mathcal{P}_2 + \mathcal{P}_2 \ar[r] & \mathcal{P}_1 \ar[r] & 0 &
}
\end{equation}
Similarly we can easily show that the relation for $\tilde{Q}_i$ is homotopic to zero due to
this diagram.
\begin{equation}
\xymatrix{
0 \ar[r] & \mathcal{P}_2 +  \mathcal{P}_2 \ar[d] \ar[r]^{(B_1,B_2)} & \mathcal{P}_1 \ar[r] \ar[d] \ar[dl]|{Q_i} & 0 & \\
0 \ar[r] & \mathcal{P}_0 \ar[r] &  0 & &
}
\end{equation}

The relation for $\tilde{A}_i$ is less trivial. By using the relations (\ref{eq:oldrel}) coming from the
old superpotential $W$, we can show that the relation for $\tilde{A}_1$ gives

\begin{equation}
\tilde{C}_1 \tilde{B}_1  - \tilde{B}_1 \tilde{C}_2 + \tilde{Q}_1 [Q_1 B_1] +\tilde{Q}_2 [Q_2 B_1] =
\binom
{C_2A_2B_1 -A_2B_2C_2 + P_1 Q_1 B_1}{-C_2A_2B_1 + A_2 B_1C_2 + P_1 Q_2 B_1 } = \binom{0}{0}
\end{equation}

The mutation from affine $A_1$ quiver with $k$ framing arrows to $k+1$ arrows is simply a straightforward
generalization of the computation.

\section{Affine $A_n$ Quiver with Framing}
\label{append:D}

The mutation machinery can be generalized to framed affine $A_n$
quiver, which is a $(n+1)$-polygon, with the framing node in the
center: \be{\xymatrix{&&\bullet\ar@(ul,ur)[]\ar[dd]
&&\\\bullet\ar@(dl,ul)[]\ar
@{<->}[urr]\ar@{<->}[dd]&&&&\bullet\ar@(ur,dr)[]\ar@{<->}[ull]\ar@{<->}[dd]\\&&\bullet
&&\\\bullet\ar@(dl,ul)[]\ar@{<->}[drr]&&&&\bullet\ar@(ur,dr)[]\ar@{<->}[dll]\\&&\bullet\ar@(dr,dl)[]&&}}\ee

One can do a chain of mutation clockwise starting from the top node
and there is an interesting pattern which is similar to affine $A_1$
case.

Schematically the mutated quiver after one step looks like:
\be{\xymatrix{&&\bullet\ar@(ul,ur)[] &&\\\bullet\ar@(dl,ul)[]\ar
@{<->}[urr]\ar@{<->}[dd]\ar[drr]&&&&\bullet\ar@(ur,dr)[]\ar@{<->}[ull]\ar@{<->}[dd]\ar[dll]\\&&\bullet\ar[uu]
&&\\\bullet\ar@(dl,ul)[]\ar@{<->}[drr]&&&&\bullet\ar@(ur,dr)[]\ar@{<->}[dll]\\&&\bullet\ar@(dr,dl)[]&&}}\ee

One can also write down the new superpotential by adding meson
coupling and integrate out massive fields.

If one keeps doing mutation clockwise, the upper right triangle will
rotate and after one cycle it becomes (notice the change in the number of framing arrows):
\be{\xymatrix{&&\bullet\ar@(ul,ur)[] &&\\\bullet\ar@(dl,ul)[]\ar
@{<->}[urr]\ar@{<->}[dd]\ar@{->>}[drr]&&&&\bullet\ar@(ur,dr)[]\ar@{<->}[ull]\ar@{<->}[dd]\\&&\bullet\ar[dll]
&&\\\bullet\ar@(dl,ul)[]\ar@{<->}[drr]&&&&\bullet\ar@(ur,dr)[]\ar@{<->}[dll]\\&&\bullet\ar@(dr,dl)[]&&}}\ee

\medskip
\medskip

The reason that this process is interesting is that the partition
function of cyclic representations of mutated quiver corresponds to
the partition function of the original quiver in some other chamber.
In other words this chain of mutation can be thought of as a series
of wall-crossing along certain directions.
\section{Deformation Complex}
\label{append:E}

For non-generic deformation $W(C_i)=\frac{1}{n+1}C_i^{n+1}$, there
are two toric actions under which the algebra is invariant. It is
conjectured that fixed points under these toric actions are isolated
and are classified by filtration of finite pyramid partition of
conifold quiver. One may compute the virtual Euler character by
using deformation complex to compute the local contributions of
these fixed points. One of the toric action will rescale the
superpotential, so the equivariant deformation complex is not
self-dual. Nevertheless due to compactness of the moduli space, one
can still get an integer weight in the end.

Under $(t,s) \in (\mathbb{C}^*)^2$, various fields
\be(A_1,A_2,B_1,B_2,C_1,C_2,P_1,P_2,\dots, P_k, Q_1,Q_2, \dots, Q_{k+1})\ee
have weights
\be{
(  \underbrace{ t_1^{-1}t_2^n,t_1 t_2^n }_{A_1, A_2}, \underbrace{t_1,t_1^{-1}}_{B_1, B_2},\underbrace{t_2,t_2}_{C_1, C_2},
\underbrace{1,t_1^2,t_1^4,\dots}_{P_1,...,P_k}, \underbrace{t_1 t_2, t_1^{-1} t_2, t_1^{-3} t_2 ,\dots}_{Q_1,...,Q_{k+1}})}\ee

Superpotential has weight $t_2^{n+1}$. One may write down the
equivariant deformation complex as follows:
\begin{equation}
\label{eq:complex1}
\xymatrix{0 \ar[r] & \mathcal{T}_0 \ar[r]^{d_1} & \mathcal{T}_1  \ar[r]^{d_2} & \mathcal{T}_2
\ar[r]^{d_3} & \mathcal{T}_3 \ar[r] & 0 }
\end{equation}

\begin{eqnarray}
\label{eq:complex2}
\mathcal{T}_0 &=& \text{End} (V_1) \oplus \text{End}(V_2) \\
\mathcal{T}_1 &=& \underbrace{\text{Hom}(V_1,V_1\otimes t_2)}_{C_1}\oplus \underbrace{\text{Hom}(V_2,V_2\otimes t_2)}_{C_2}\oplus
\underbrace{\text{Hom}(V_1,V_2\otimes(t_1^{-1}t_2^n\oplus t_1 t_2^n))}_{A_1, A_2} \nonumber \\
&&\oplus \underbrace{\text{Hom}(V_2,V_1\otimes(t_1\oplus t_1^{-1}))}_{B_1,B_2} \oplus
\underbrace{\text{Hom} (V_2 , \mathbb{C} \otimes (1\oplus t_1^{2}\oplus t_1^{4} \oplus \dots))}_{P_1,...,P_k} \nonumber \\
&& \oplus \underbrace{\text{Hom}(\mathbb{C},V_1\otimes
t_1 t_2 (1\oplus t_1^{-2} \oplus t_1^{-4} \oplus \dots))}_{Q_1,..,Q_{k+1}} \\
\mathcal{T}_2 &=& \mathcal{T}_1^*\otimes t_2^{n+1} \\
\mathcal{T}_3 &=& \mathcal{T}_0^*\otimes t_2^{n+1}
\end{eqnarray}

where we tensor each vector space with some representation of $( \mathbb{C}^* )^2$ of certain
charge to make the complex equivariant.

We now specialize to the case of $k=1$, namely the quiver as follows. The general $k$ case
is a straightforward generalization.

\begin{equation}
\label{eq:quiver2}
\xymatrix{
& \bullet 0 \ar[ddl]_{P_1} &  \\
& & \\
\bullet 1 \ar@(ul,dl)[]|{C_1}
\ar@/^/@{->>}[rr]^{A_1, A_2} && \bullet 2  \ar@{->>}[uul]_{Q_1,Q_2}  \ar@(ur,dr)[]|{C_2} \ar@/^/@{->>}[ll]^{B_1, B_2}  \\
}
\end{equation}
\begin{equation}
W= w(C_1)-w(C_2) + C_1 ( B_1 A_1 + B_2 A_2) -C_2 ( A_1 B_1 +A_2 B_2) +A_1 P_1 Q_1 + A_2 P_1 Q_2
\end{equation}

Note that this quiver (\ref{eq:quiver2}) is basically the same as (\ref{eq:quiver1}) with $A_i$ and $B_i$ swapped.

We use the capital letter ($A_i, B_i, C_i, P_1, Q_i$, $i=1,2$) to represent the fields, satisfying the F-term and D-term condition. Namely
it represents a point in the quiver moduli space. The small letters ($a_i, a_i, c_i, p_1, q_i$) are the fluctuations around this points.
$d_1$ is given by the gauge transformation:

\be
d_1\binom{\gamma_1}{\gamma_2}=\left(\begin{array}{cc}\gamma_1C_1-C_1\gamma_1\\\gamma_2C_2-C_2\gamma_2\\\gamma_2
A_i-A_i\gamma_1
\\\gamma_1
B_i-B_i \gamma_2\\\gamma_1 P_1 \\-Q_i\gamma_2\end{array}\right)
\ee

We linearize the F-term relations around the point  $(A_i, B_i, C_i, P_1, Q_i)$ to obtain $d_2$:

\be
d_2\left(\begin{array}{cc}c_1\\c_2\\a_i\\b_i\\p_1\\q_i\end{array}\right)=
\left(\begin{array}{cc}c_1C_1^{n-1}+C_1c_1C_1^{n-2}+\dots+C_1^{n-1}c_1+b_1A_1+B_1a_1+b_2A_2+B_2a_2\\
c_2C_2^{n-1}+\dots+C_2^{n-1}c_2+A_1b_1+a_1B_1+A_2b_2+a_2B_2\\
p_1 Q_i+ P_1 q_i+c_1B_i+C_1 b_i-b_i C_2-B_ic_2\\
a_iC_1+A_ic_1-c_2A_i-C_2a_i\\
q_1 A_1+Q_1 a_1+ q_2 A+2+Q_2a_2\\
A_i p_1 +a_i P_1
\end{array}\right)
\ee

One needs to play a bit with the F-term relations in quiver (\ref{eq:quiver2})
to find out the \textit{relations of F-term relations}.

\be
d_3\left(\begin{array}{cc} \delta_1 \\ \delta_2 \\ \alpha_i \\ \beta_i \\ \kappa \\ \pi_i \end{array}\right)=
\left(\begin{array}{cc} \delta_1 C_1- C_1 \delta_1 - B_1 \beta_1 - B_2 \beta_2 + \alpha_1 A_1+ \alpha_2 A_2 + P_1 \kappa \\
\delta_2 C_2-C_2 \delta_2 + A_1 \alpha_1 + A_2 \alpha_2-\beta_1 B_1-\beta_2 B_2-\pi_1 Q_1-\pi_2 Q_2\end{array}\right)
\ee

It is a simple exercise to show that $d_2 \circ d_1 = d_3 \circ d_2 = 0$

For a dimension vector $(v_1,v_2)$ or equivalently D-brane charge
vector $(p,q)$, which is related by a $SL(2,\mathbb{Z})$
transformation depending on $k$, one first lists all possible
partitions (filtered pyramid partitions) and then decomposes $V_1$ and $V_2$
in terms of $(\mathbb{C}^*)^2$ representation.

Using this $V_1$ and $V_2$ decomposition we can furthur decompose
the linear spaces in (\ref{eq:complex1}) and
write
\be
\mathcal{T}_i =  \sum k_{i,(n_1,n_2)} t_1^{n_1} t_2^{n_2} .
\ee

We then associate a polynomial $P_i$ with $\mathcal{T}_i$.

\be
\mathcal{T}_i = \sum k_{i,(n_1,n_2)} t_1^{n_1} t_2^{n_2} \to P_i(\alpha_1, \alpha_2)=\prod_{(n_1,n_2)} (n_1 \alpha_1 + n_2 \alpha_2)^{k_{i,(n_1,n_2)}} \ee

After taking the alternating ratio one
gets \be{D(p,q)=\sum\frac{P_0P_2}{P_1P_3}}\ee where the sum is over
all the fix points of dimension vector $(v_1,v_2)$.
In fact we can also compute
\be
\mathcal{T} =  \mathcal{T}_0-\mathcal{T}_1+\mathcal{T}_2-\mathcal{T}_3= \sum k_{(n_1,n_2)} t_1^{n_1} t_2^{n_2} .
\ee
all at once and find the polynomial $P=\frac{P_0P_2}{P_1P_3}$ associated with $\mathcal{T}$.

Notice that the deformation complex has the property of being
twisted self-dual, which means that if
\be{P_1P_3(\alpha_1,\alpha_2)=\prod_{(n_1,n_2)}(n_1\alpha_1+n_2\alpha_2)}\ee
then
\be{P_0P_2(\alpha_1,\alpha_2)=\prod_{(n_1,n_2)}(-n_1\alpha_1-n_2\alpha_2+(n+1)\alpha_2)}\ee
for each fix point, the local contribution is:
\be{\prod_{(n_1,n_2)}\frac{-n_1\alpha_1-n_2\alpha_2+(n+1)\alpha_2}{n_1\alpha_1+n_2\alpha_2}}\ee
Unlike conifold, these linear factors do no cancel in pairs and
contribute a $(-1)$. One may want to set $\alpha_2=0$ when
$n_1\neq0$, which still gives a $(-1)$ factor. When $n_1=0$ one gets
a non-trivial weight $\frac{-n_2+n+1}{n_2}$. To conclude each local
contribution looks like: \be{\prod_{n_1\neq
0,n_2}(-1)\prod_{n_1=0,n_2}\frac{-n_2+n+1}{n_2}}\ee

$(n_1,n_2)=(0,0)$ or $(0,n+1)$ needs more attention, to get a
non-zero weight, there should be equal number of $(0,0)$ and
$(0,n+1)$. In fact it is possible there is extra $(0,n+1)$ which
means that weight is zero.

Although it has been checked in many examples that the classification of the fixed points is correct.
it remains to come up with a systematic combinatoric statement.

\end{document}